\documentstyle[prd,aps,psfig]{revtex}
\bibstyle{unsrt}

\tighten
\begin{document}
\draft
%
%
\twocolumn[\hsize\textwidth\columnwidth\hsize\csname
@twocolumnfalse\endcsname

\date{\today}
\author{Graham Vincent, Nuno D. Antunes and Mark Hindmarsh}
\title{Numerical simulations of string networks in the
Abelian-Higgs model}
\address{Centre for Theoretical Physics\\
University of Sussex\\
Brighton BN1 9QJ\\
U.K.}
\preprint{SUSX-TH-97-015, hep-ph/9708427}

\maketitle

\begin{abstract}
We present the results of a field theory simulation of networks of
strings in the Abelian Higgs model. Starting from a random initial
configuration we show that the resulting vortex tangle approaches
a self-similar regime in which the length density of lines of zeros
of $\phi$ reduces as $t^{-2}$. We demonstrate that the network
loses energy directly into scalar and gauge radiation.
These results support a recent claim that
particle production, and not gravitational radiation, is the dominant 
energy loss mechanism for cosmic strings. This means that
cosmic strings in Grand Unified Theories
are severely constrained by high energy cosmic ray fluxes:
either they are ruled out, or an implausibly small fraction
of their energy ends up in quarks and leptons.
\end{abstract}
\pacs{PACS numbers: 98.80.Cq 11.27.+d \hfill 
Preprint: SUSX-TH-97-015, hep-ph/9708427}

\vskip2pc]

In recent years, a sustained effort has gone into understanding
the formation and evolution of networks of cosmic string, principally
to provide a mechanism for seeding gravitational collapse 
\cite{HindKib94,ShelVil}.

The picture that emerges from lattice simulations 
of string formation is a network consisting of a
small number of horizon crossing self avoiding random walks,
together with a scale invariant distribution of loops \cite{Lattice}.
The subsequent evolution is driven by a tension in the strings
causing them to straighten out. When two lengths of string pass through 
each other,
they may {\it intercommute}, that is exchange partners.
This allows the production of loops which
can decay through gravitational radiation or particle
production, depending on their size.

A number of numerical studies of network evolution
have been carried out using the Nambu-Goto approximation of 
the string as a 1-dimensional object with a tension $\mu$.
In particular we will refer later to work by Allen and Shellard (AS)
and Bennett and Bouchet (BB), both reported in \cite{FRWCodes}.
A consensus emerged from these studies that the network will
relax into a {\it scaling} regime, where the large scale features of
the network (the inter-string distance $\xi_p$ and the 
step length $\bar\xi_p$) grow with the horizon in proportion
to $t$.
On scales less than $\bar\xi_p$, there was evidence for
a fractal substructure covering a range from the resolution scale to 
$\bar\xi_p$, which seems to be the result of
kinks left behind by the production of loops.
The loop production function itself did {\it not}
scale: the distribution of loop sizes reported by BB
is peaked around a cut-off introduced by hand 
into the simulation to ensure that even the smallest 
loops are approximated by a reasonable number of points.

It has been thought that the build-up 
of small scale structure will only be stopped by back-reaction
from the string's own gravitational field when the intermediate fractal
extends to scales of order $G\mu\xi_p$, where $G$ is Newton's constant. 
Then the small scale structure
and loop production will scale with the horizon. Although much smaller 
than 
$\xi_p$, loops produced at this scale are still vastly bigger than the 
string core and will decay through gravitational radiation.

This picture has some support from a detailed analytical study by
Austin, Copeland and Kibble \cite{ACK} (ACK) which deals in part with
length scales $\xi_p$, $\bar\xi_p$ and a scale $\zeta$, which can be 
interpreted as an angle-weighted average distance between kinks on 
the string. However, the ACK analysis relies
on a number of unknown parameters,
and for some parameter ranges small scale structure is absent.

In \cite{VHS}, Vincent, Hindmarsh and Sakellariadou (VHS) 
suggested a different picture 
based on results from Minkowski space Nambu-Goto simulations using the 
Smith-Vilenkin algorithm \cite{SmithVil:alg}.
This algorithm is exact for string points defined on a lattice, allowing
easy detection of intercommutation events.
When loop production is unrestricted
(up to the lattice spacing) they found that small scale structure
disappeared, although loop production 
continued to occur at the lattice spacing. 
They could only recover the small scale structure seen in 
\cite{FRWCodes} 
by artificially restricting loop production
with a minimum loop size greater than the lattice spacing. 
The suggestion is that the small scale structure
seen in other simulations is an artifact of a minimum loop size and that 
loop production will occur at the smallest physical scale: the string 
width.

This has interesting consequences for energy loss from string
networks. Loops formed at this size will decay into particles rather
than gravitational radiation and could give a detectable flux 
\cite{BhattRana}.
If particle production provides 
the dominant energy loss
channel, then GUT theories with strings are heavily constrained.

In this letter, we present the results of a series of field theory 
simulations
of networks of Abelian Higgs vortices. They support the 
results of Vincent {\it et\,al\,}.
If loops form, they form at the string width scale
and promptly collapse. However, most of the string
energy goes directly into oscillations 
of the field - radiation. We measure little small scale
structure in the network. Furthermore, we find that
the network scales with a scaling density consistent 
with the Smith-Vilenkin simulations,
implying that the latter simulations do not unduly 
exaggerate energy loss
as a lattice effect, a possibility suggested in \cite{Alb90}.

The simplest gauge strings are contained in the
Abelian-Higgs model, which has a Lagrangian
\begin{equation}
{\mathcal{L}} = (D_{\mu}\phi)^\dagger(D^{\mu}\phi) 
            - \frac{1}{4}F_{\mu\nu}F^{\mu\nu}
            - \frac{\lambda}{4}(|\phi|^2-\sigma^2)^2,
\label {ActionAH}
\end{equation}
where $\phi$ is a complex scalar field, gauged by a U(1) 
vector potential with a covariant derivative
$D_{\mu}=\partial_{\mu}-ieA_{\mu}$. $F_{\mu\nu}$ is the field
strength tensor
$\partial_{\mu}A_{\nu}-\partial_{\nu}A_{\mu}$.
To model the system described by Eq.\ \ref{ActionAH} 
we use techniques from hamiltonian lattice
gauge theory \cite{LGT} (and references therein).

An attractive feature of this formalism is 
that the Hamiltonian respects the discrete
version of the $U(1)$ gauge transformations 
and consequently if Gauss's law is true
initially, then it is true for all later times. 
Significant violations of Gauss's law
are obtained if instead one uses a numerical scheme based on 
finite-differencing
the Euler-Lagrange equations for $\phi$ and $A_i$.

The equations of motion derived from the Hamiltonian are evolved by 
discretising time
with $\Delta t = ra$ ($r\ll 1$) and using the leapfrog method of 
updating
field values on even time steps and the conjugate momentum on odd time 
steps.
In keeping with the systems we are trying to model, we create initial 
conditions 
by allowing an energetic configuration of fields to dissipate energy 
until
$\phi$ is close to the vacuum everywhere except near the string cores.
We are not attempting to model the formation process, rather we wish to 
create a reasonable random 
network of flux vortices for subsequent evolution.

The simulation proceeds as follows. We generate a Gaussian random 
$k$-space 
configuration of each component of $\phi$
with a $k$-dependent Gaussian probability distribution 
with variance $\sigma_{\bf k}=    {\sqrt{m\;b}/{({\bf k}^2 + m^2})}$.
Samples drawn from this distribution are then Fourier transformed to 
$x$-space to generate a starting configuration, which will be 
uncorrelated
on scales greater than $m^{-1}$.
By varying the two paramenters $m$ and $b$ it is possible to control 
both
the
average amplitude and the correlation length of the initial 
configuration
(this is in turn related to the initial defect density after the 
dissipation
period). This then forms the initial conditions
to be dissipated along with 
$\dot\phi=\theta^i=E^i=0$. Our dissipation scheme is to add the 
gauge invariant terms
$\eta\pi_x$ and $\eta E_x^i$ to the equations of motion for $\phi$ 
and $\theta^i$.
This ensures that Gauss's law is preserved by the dissipation, 
but changes the charge-current 
continuity equation so that
$
\dot\rho + \nabla\cdot j = -\eta\rho.
$
We have found that the most efficient dissipation is when the charge
grows from $0$ to $e$ during the dissipation process. This 
allows the early formation of essentially
global vortices, into which the magnetic flux relaxes as $e$ is 
increased.

Once the network has formed with $\langle{|\phi|\rangle}\approx 1$, 
we evolve the network
with the dissipation turned off for half the box light-crossing 
time. During the
evolution energy is conserved to within $\approx 0.3\%$.

A reasonable resolution of the string core is obviously an
important requirement of the simulation. 
Traditionally the core is defined by the inverse masses 
of the field, so for $e=1$, $\lambda=2$ string, $d_c=1$. However, the
fields depart appreciably from the vacuum over a larger distance, 
about 4 units.

We have performed a series of simulations on lattices of $224^3$,
$304^3$ and $336^3$ points, lattice spacings of between 
$0.25$ and $0.75$, and with $r=0.15$.
We vary the parameter $m$ in the variance of the 
field distribution to control initial defect density.
In this letter we keep $\lambda/2e^2 =1$.

Due to the size of the simulations we are
unable to achieve the same level of statistical significance obtained
using Nambu-Goto codes, although there are some revealing qualitative
results. 
\begin{figure}[b!]
\centerline{\psfig{file=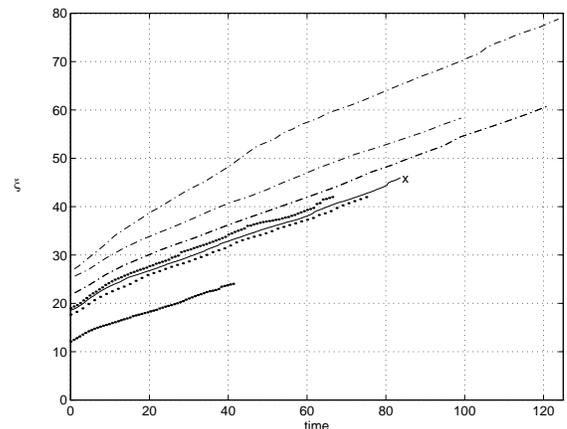,width=2.9in}}
\caption{Plots of $\xi_p$ for a series of $336^3$
simulations with different lattice spacings.
From top to botttom $a=$0.75, 0.65, 0.75, 0.4,
0.5, 0.45 and 0.25. In the initial gaussian
distribution the parameter $m$ is varied to give
the different initial correlation lengths.
$\xi$ is given in units of the inverse
scalar mass $m_s^{-1}$.}
\label{xis}
\end{figure}
The gauge-invariant zero of the $\phi$
field provides a simple way of analysing the
network of vortices. A string passes through a 
lattice cell face if the winding around
the four corners is non-zero. By starting in any 
lattice cell with a string in it, 
the string can be followed around the lattice until 
it returns to the starting point.
The distribution of box-crossing string and loops can 
then be analysed in much the
same way as for Nambu-Goto strings \cite{FRWCodes,VHS}. 
In particular, we are interested in
the behaviour of the length scale $\xi_p$, roughly the 
inter-string distance, defined as
$
\xi_p={V / L^2_{\xi_p}}
$,
where $L_{\xi_p}$ is the total {\it physical} length of string 
above length $\xi_p$ (not the more usual invariant length). 
In practise, we calculate
the length of string by smoothing over the string 
core width $d_c$. Scaling occurs if
$\xi_p$ grows with $t$, although the actual value of 
$x_p$ in the scaling density
$
\rho_{\xi_p}={\mu / \xi_p^2}={\mu / x_p^2 t^2}
$,
depends on the efficiency of the energy loss mechanism.
Fig.\ \ref{xis} shows the function $\xi_p$ for a sample set of runs
with different simulation parameters. Although with small dynamic 
ranges,
one can never be sure that $\xi_p$ is 
approaching scaling or just on a slow transient, 
the scaling values for $x$ all appear to be in the range 
$0.27-0.34$, which is in agreement with Smith-Vilenkin simulations
with maximum loop production where $x_p=0.27\pm 0.05$. 
This value is obtained by converting
the invariant length $L_{\xi}$ used to define $\xi$ in 
\cite{VHS} to physical 
length  $L_{\xi_p}$, if $\bf X' (\sigma)$ is the string 
tangent vector, then the physical
length is just $L_{\xi_p}=\int d\sigma | \bf X'|$.
In the latter half of the simulation runs,
these plots are extremely linear: the exponent of
$t$ in $\xi\propto t^p$ is $p=1.002\pm 0.03$.

\begin{figure}[!]
\centerline{\psfig{file=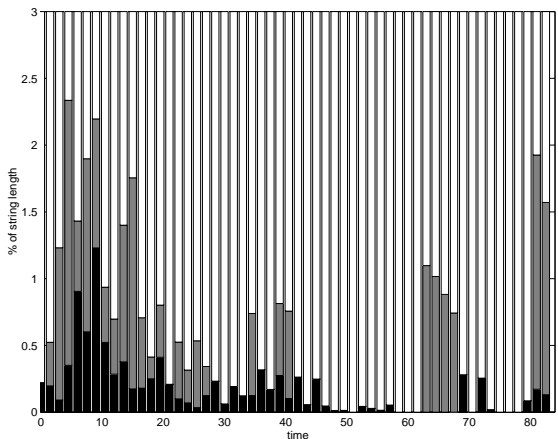,width=2.9in}}
\caption{Proportion of string length intermediate loops
(grey) and core loops (black). 
The long string (white) contribution makes up $100\%$.
The lattice spacing is $a=0.35$.}
\label{fig:loops}
\end{figure}

In \cite{VHS}, VHS argued that the string network 
is be dominated by 
processes on the smallest physical scale, either radiation
or the production of loops at the string width.
In all our runs performed so far,
we find that this is true. When analysing
the distribution of string formed by zeros of $\phi$,
we consider three classes: long string ($L>\xi_p$), intermediate
loops ($\pi d_c<L<\xi_p$)
and core loops ($L<\pi d_c$). Fig.\ \ref{fig:loops} shows
a typical run for with a proportionate break down into types of loop 
over time.

Over the first part of the simulation we see an overall 
decrease in the percentage of
string in loops of both types. Over the 
latter part of the simulation we see three sharp increases in $\%$ 
of intermediate loops
at $t=37$, $t=61$ and $t=81$. In all three cases, this is caused by a 
single
large loop (with $L>\xi_p$) decaying to below the $\xi_p$ threshold.
Although these events constitutes a loss to the long string network, the
decay channel is radiation and {\it not} an
intercommutation event producing an intermediate loop. Note that aside 
from
these loops, in the latter half of the simulation,
{\it all the string length is in long string (together with
a small population of core loops), and the network continues to scale},
despite 
some minor fluctuations in $\xi_p$  
corresponding to the shrinking large loops (see curve marked `x' in 
Fig.\ \ref{xis}).
We infer that all energy loss occurs ultimately through radiation
and that it is very efficient at scaling the network.

To back up our claim that radiative processes are key
to network evolution, we examined sinusoidal standing waves 
with wavelength $\lambda$ and initial amplitude $A$
where $A=\lambda/2$. The necessary scaling behaviour for
the energy density in a network can be seen in the
radiation from this system (incidentally one without intercommutation).
Consider that a string network can be thought of
as a length $\xi$ in a volume $\xi^3$, giving a density 
$\rho=\mu/\xi^2$ and $\dot\rho=-2\mu\dot\xi/\xi^3$. For a scaling 
network,
$\dot\rho\xi^3=$ constant. We studied a series of standing waves
with increasing wavlength $\lambda$ and measured the energy density
$\rho(t)$ in a sub-volume $V_{\lambda}\sim\lambda^3$ around the string.
We found that $\rho$ decays fairly linearly as 
the radiation from the string begins to leave $V_{\lambda}$,
and then tails off as $A$ decreases significantly.
We measure $\dot\rho$ during the linear region
when $A\approx\lambda/2$. We find that
$\dot\rho\lambda^3$ is constant to within 10 percent over a range of 
standing waves with $\lambda$ from 20 to 180, two orders of magnitude
bigger than the string width. We also note that the pattern of 
oscillations 
around the standing wave remains
substantially unchanged by an increase in the lattice spacing:
we believe that any differences can be accounted for by the resulting 
decrease
in resolution. The fact that the scaling density in the network
simulation does not depend on the lattice spacing
is also evidence that the radiative process is not a lattice artifact.
\begin{figure}[!]
\centerline{\psfig{file=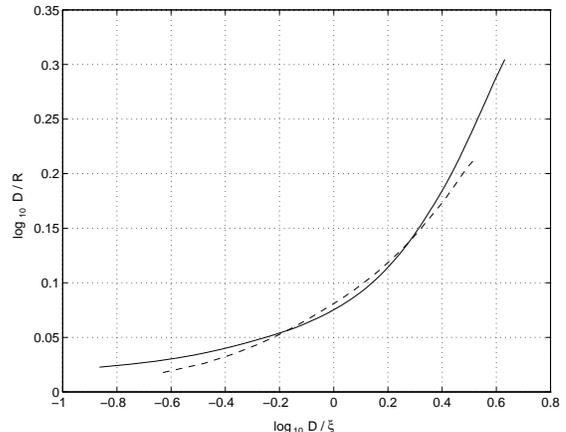,width=2.9in}}
\caption{The fractal dimension of the long string at two times
late in the evolution for $B=336^3$ and $a=0.75$ (solid), $a=0.25$ 
(dashed).
The curves do not coincide, which may be due to difference in 
dynamic range in the two simulations.}
\label{fig2}
\end{figure}
It is claimed in \cite{VHS} that if loop production is unrestricted,
small scale structure will disappear.  Small scale structure has been 
studied
in two ways. Firstly, there is a length scale $\zeta$ 
mentioned in the introduction. Secondly, the relation between
end-to-end distance $R$ and distance along the string $D$, 
$
(D / R ) \propto (D / \xi_p ) ^n,
$
defines a scale-dependent fractal dimension $n$.
If strings are smooth on small 
scales and brownian on large scales,
$n$ should interpolate from $\sim 0$ to $\sim 0.5$.
BB and AS observed that over a significant fraction of the range
from the resolution scale to $\xi$, $n$ is constant.
VHS showed from the behaviour of $\zeta$ that the region of
constant $n$ did reflect the existence of small scale structure:
when loop production is unrestricted the region of constant $n$ is 
absent
and $\zeta \sim \xi$; when loop
production is restricted a region of constant $n$ emerges 
and $\zeta \ll \xi$.

Fig.\ \ref{fig2} shows that for the flux vortices $n$ varies 
smoothly from $\sim 0.01$
to $\sim 0.5$. The absence of the transient is in good 
qualitative agreement with the Nambu-Goto simulations with 
$\zeta\sim\xi$,
and we can argue from this that gravitational radiation ({\rm gr})
is irrelevant to our claimed particle production ({\rm pp}).
Energy loss processes can be described by an efficiency parameter $c$ 
in the rate equation $\dot\rho=-c\rho/\xi$. For a scaling network
with $\zeta=\epsilon\xi$, $c_{{\rm gr}}=\Gamma G\mu/\epsilon$
where $\Gamma\sim 10^1-10^2$. If there is no small scale structure,
$\epsilon\sim 1$ and, for GUT strings, $c_{{\rm gr}}=10^{-4}-10^{-5}$.
This is to be compared with the overall energy loss
through radiation in our simulations which gives $c_{{\rm pp}}=0.6$.
It is only when there is considerable small scale structure
($\epsilon\sim 10^{-4}$) that gravitational radiation becomes
significant.

We believe that our results support the claim of VHS that
the dominant energy loss mechanism of a cosmic string network
is particle production. Our field theory simulations show 
that the string network
approaches self-similarity, as $\xi_p \propto t$, 
with a constant of proportionality $x_p$
consistent with Nambu-Goto codes which simulate 
string objects directly. We provide
evidence that the long string network loses energy 
directly into oscillations of the field, which 
is the classical counterpart 
of particle production, or into loops of order the string width
which quickly decay into particles.
Previous calculations on particle production underestimated 
its importance by using perturbative calculations 
with light particles \cite{PertRad}. We see non-perturbative
radiation at the string energy scale.

This represents a radical divergence from the traditional cosmic 
string scenario, which held that the loops 
are produced at scales much greater than
the string width and subsequently decay by 
gravitational radiation. It implies that strong constraints
come from the flux of ultra high energy (UHE) cosmic rays.
Bhattacharjee and Rana \cite{BhattRana} estimated that no more 
than about $10^{-3}$
of the energy of a GUT-scale string network can be injected as
grand unified $X$ particles (assumed to decay into quarks and leptons).
More recent calculations \cite{ProtStan} limit the current energy
injection rate $Q_0$
of particles of mass $m_X$ to approximately 
$Q_0 \lesssim 2 \times 10^{-34} (m_X/10^{15}\,{\rm GeV})^{0.44}
{\rm\, ergs\, s}^{-1} {\rm cm}^{-3}$ (this
formula is based on an eyeball fit to their most conservative bound
in Fig. 2 \cite{ProtStan}). Sigl 
{\it et\,al} \cite{Sigl} obtain somewhat lower values for $Q_0$: 
we have been cautious
and adopted the higher value as our bound.

The energy injection rate from a scaling network is of order
$\mu/t^3\sim10^{-32}
{\rm\,ergs\,s}^{-1} {\rm cm}^{-3}\,(\mu/10^{32}\,{\rm GeV}^2$).
Taking $\mu\simeq 10 m_X^2$,
we derive $\mu\lesssim 10^{-29} f_X^{-1.3}\,{\rm GeV}^2$ or 
$G\mu \lesssim 10^{-9} f_X^{-1.3}$,
where $f_X$ is the fraction of the energy appearing as quarks and 
leptons.
Thus we conclude that GUT scale strings with $G\mu \sim 10^{-6}$ 
are ruled out: 
we regard a fraction $f_X \sim 5\times 10^{-3}$ as implausible.

We would like to thank Ray Protheroe,  Hugh Shanahan
and G\"unter Sigl for useful discussions, and Stuart Rankin 
for help with the UK-CCC facility in Cambridge. 
This research was conducted in cooperation with Silicon Graphics/Cray
Research
utilising the Origin 2000 supercomputer and supported by HEFCE and 
PPARC.
GV and MH are supported by PPARC, by studentship number 
94313367,  Advanced Fellowship number B/93/AF/1642 
and grant numbers GR/K55967 and GR/L12899. NDA is supported by
J.N.I.C.T. - {\it Programa Praxis} XXI,
under contract BD/2794/93-RM.
Partial support is also obtained from the European Commission 
under the Human Capital 
and Mobility programme, contract no. CHRX-CT94-0423 and 
from the European Science Foundation.

\end{document}